\documentclass[allclo]{FBSart}
\usepackage{amsfonts}
\usepackage{amssymb}
\usepackage{epsf}
\usepackage[dvips]{graphicx}

\title{Ab-initio calculations of four-nucleon elastic scattering}
\author{R.~Lazauskas\thanks{\textit{E-mail address:}
 lazauskas@lpsc.in2p3.fr}, J.~Carbonell\thanks{\textit{E-mail address:}
 carbonell@lpsc.in2p3.fr}.} \institute{Laboratoire de
Physique Subatomique et de Cosmologie,
        53. avenue des Martyrs, 38026 Grenoble Cedex, France}

\runningauthor{R.~Lazauskas, J.~Carbonell } \runningtitle{Full
microscopic calculations of four-nucleon elastic scattering}
\begin{document}

\maketitle
\begin{abstract}
  We present microscopic calculations of low energy scattering
observables in all possible four nucleon systems: n-$^3$H,
p-$^3$He and p-$^3$H. Results were obtained by solving
Faddeev-Yakubovski equations in configuration space, appropriately
modified to include Coulomb and three-nucleon forces.
\end{abstract}

\section{Introduction}

 Three- and four-nucleon systems are the testing ground for
studying the nuclear interaction. If modern NN potentials have
reached a very high degree of accuracy in describing the
two-nucleon data, they fail already to account for the binding
energies of the lightest nuclei. The use of three-nucleon forces
(3NF) is -- with some exceptions \cite{Doles,Gross} -- mandatory.
By adjusting its parameters one can obtain a satisfactory
description of nuclear binding energies up to A=10 \cite{Pieper}.

Apart the effect of rescaling nuclear thresholds, low energy
three-nucleon scattering observables are quite insensitive to 3NF.
Furthermore, since 3N spectra does not contain any narrow
resonances,  three-nucleon dynamics is relatively rigid once
deuteron and triton binding energies are fixed.
\begin{figure}[ht]
\begin{center}
\vspace{-0.25cm}\includegraphics[width=12.5cm]{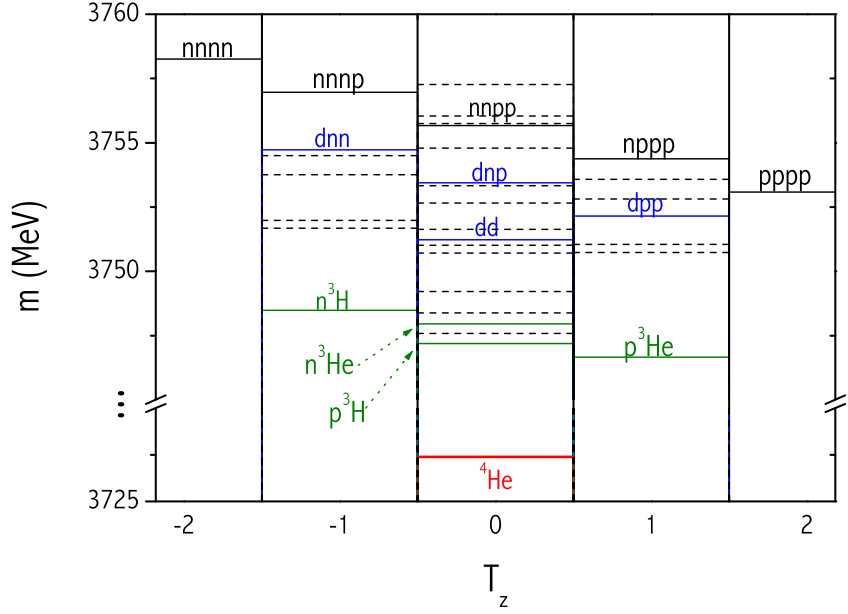}\vspace{-1.0cm}
\caption{Four nucleon energy spectra. Single lines indicate
particle thresholds, whereas dashed lines represent particle decay
unstable excited states (resonances).} \label{Fig_n4_en_chart}
\end{center}\end{figure}
 The four nucleon
continuum is a challenging few-body nuclear problem. Its interest
lies not only in the natural progression that it represents,
towards a systematic description of nuclear systems with
increasing complexity, but also in the richness of the $A=4$
nuclear chart itself (see Fig. \ref {Fig_n4_en_chart}). Four
nucleon problem, we believe, implies a qualitative jump in respect
to the $A=3$ case.

\section{Theoretical tools}

We use the Faddeev-Yakubovski (FY) equations in configuration
space to tackle the four-particle problem. However in their
original form these equations are applicable only for systems
interacting via short-range pairwise forces. An elegant way to
include 3NF into FY equations was suggested in \cite{Gloeck_3BF}
for four identical particles. These equations reads:
\begin{eqnarray}
\left(E-H_{0}-V_{12}\right)K_{12,3}^{4}&=&V_{12}(P^{+}+P^{-})\left[
(1+Q)K_{12,3}^{4}+H_{12}^{34}\right] +V_{12,3}\Psi \nonumber \\
\left( E-H_{0}-V_{12}\right) H_{12}^{34} &=&V_{12}\tilde{P}\left[
(1+Q)K_{12,3}^{4}+H_{12}^{34}\right]\label{FY1}
\end{eqnarray}
with $P^{+}$, $P^{-}$, $\tilde{P}$ and $Q$ being permutation
operators:
\begin{eqnarray}
\begin{tabular}{lll}
$P^{+}=(P^{-})^{-}=P_{23}P_{12};$ & $Q=\varepsilon P_{34}$; & $\tilde{P}%
=P_{13}P_{24}=P_{24}P_{13}$.
\end{tabular}
\end{eqnarray}
and
\[
\Psi =\left[ 1+(1+P^{+}+P^{-})\right]
Q(1+P^{+}+P^{-})K_{12,3}^{4}+(1+P^{+}+P^{-})(1+\tilde{P})H_{12}^{34}
\]
the total wave function.

Equation (\ref{FY1}) becomes however non appropriate once long
range interaction,
 in particular Coulomb, is present. In fact, FY components remain coupled even
 in far asymptotics, thus making the implementation of correct boundary conditions
hardly possible. In order to circumvent this problem we have split
Coulomb potential ($V_{C}$) into long ($V_{C}^{(l)}$) and  short
($V_{C}^{(sh)}$) range parts by means of some arbitrary cut-off
function $\chi (x,y,z)$, similarly as it was done by Merkuriev for
3-body equations \cite{Merk_AP130}:
\begin{equation}
V_{C}(x) =V_{C}^{(sh)}(x,y,z)+V_{C}^{(l)}(x,y,z); \qquad
\begin{tabular}{l}
$V_{C}^{(sh)}(x,y,z)=\chi (x,y,z)V_{C}(x)$ \\
$V_{C}^{(l)}(x,y,z)=\left[ 1-\chi (x,y,z)\right] V_{C}(x)$%
\end{tabular}
\end{equation}
$\chi (x,y,z)$ is a smooth function which equals 1 for small  $x$
values or when $x\ll (y,z)$ and vanishes for $x\gg (y,z).$  Then
the 4-body FY equations can be rewritten in the form:
\begin{small}
\begin{eqnarray}
(E-H_{0}-V_{12}^{(sh)}-\sum V_{C}^{(l)})K_{12,3}^{4}
&=&V_{12}^{(sh)}(P^{+}+P^{-})\left[
(1+Q)K_{12,3}^{4}+H_{12}^{34}\right]+V_{12,3}\Psi \nonumber \\
(E-H_{0}-V_{12}^{(sh)}-\sum V_{C}^{(l)})H_{12}^{34}&=&V_{12}^{(sh)}\tilde{P}%
\left[ (1+Q)K_{12,3}^{4}+H_{12}^{34}\right]\label{MFY_eq}
\end{eqnarray}\end{small}
We use for short the following notations:
\begin{equation}
V_{12}^{(sh)}=\left( V_{(N)}+V_{C}^{(sh)}\right) _{12}.
\end{equation}

In this form FY equations become assymptoticaly uncoupled and
appropriate boundary conditions can be easily implemented.

Equations (\ref{MFY_eq}) are solved by making partial wave
decomposition of amplitudes $K_{12,3}^{4}$ and $H_{12}^{34}$:

\begin{eqnarray}
K_{i}(\vec{x}_{i},\vec{y}_{i},\vec{z}_{i}) &=&\sum_{LST}\frac{\mathcal{K}%
_{i}^{LST}(x_{i},y_{i},z_{i})}{x_{i}y_{i}z_{i}}\left[ L(\hat{x}_{i},\hat{y}%
_{i},\hat{z}_{i})\otimes S_{i}\otimes T_{i}\right]  \\
H_{i}(\vec{x}_{i},\vec{y}_{i},\vec{z}_{i}) &=&\sum_{LST}\frac{\mathcal{H}%
_{i}^{LST}(x_{i},y_{i},z_{i})}{x_{i}y_{i}z_{i}}\left[ L(\hat{x}_{i},\hat{y}%
_{i},\hat{z}_{i})\otimes S_{i}\otimes T_{i}\right]
\end{eqnarray}

The partial components $\mathcal{K}_{i}^{LST}$ and $\mathcal{H}_{i}^{LST}$%
are expanded in the basis of three-dimensional splines. Thus
integro-differential equation (\ref{MFY_eq}) is converted into
system of linear equations. For detailed discussion on the
equations and method used one can refer to \cite{Thesis}.

\section{Results}
\subsection{n-$^3$H elastic scattering}

The elastic scattering of neutrons on tritium is the simplest 4N
reaction. Treatment of this system doesn't require calculation of
the costly integrals due to the Coulomb terms in eq.
(\ref{MFY_eq}). Nevertheless this system contains two, spin
degenerated, narrow resonances at low energy ($E_{cm} \approx 3$
MeV). Ability of nuclear interaction models to describe scattering
cross sections in this resonance region is still an open question
\cite{Fonseca,Carbonell1}.

\begin{table}[tbh]
\begin{tabular}{c||l|l||l|l||l|l|}
\cline{2-7} & \multicolumn{2}{|c||}{MT I-III} &
\multicolumn{2}{|c||}{Av. 14} & \multicolumn{2}{|c|}{Av. 18+UIX}
\\ \cline{2-7} \multicolumn{1}{l||}{} & J$^{\pi }=0^{+}$ & J$^{\pi
}=1^{+}$ & J$^{\pi
}=0^{+}$ & J$^{\pi }=1^{+}$ & J$^{\pi }=0^{+}$ & J$^{\pi }=1^{+}$ \\
\hline\hline \multicolumn{1}{|l||}{n-$^{3}$H} & 4.10 & 3.63 & 4.28
& 3.81 & 4.04 & 3.60
\\ \hline
\multicolumn{1}{|l||}{p-$^{3}$He} & 11.5 & 9.20 & 12.7 & - & 11.3 & - \\
\hline \multicolumn{1}{|l||}{p-$^{3}$H} & -63.1 & 5.50 & -13.9 &
5.77 & -16.5 & 5.39
\\ \hline
\end{tabular}
\caption{4N scattering lengths calculated using different
interaction models. } \label{table:1}
\end{table}

\begin{figure}[ht]
\vspace{-0.25cm}
\par
\begin{center}
\includegraphics[width=12.5cm]{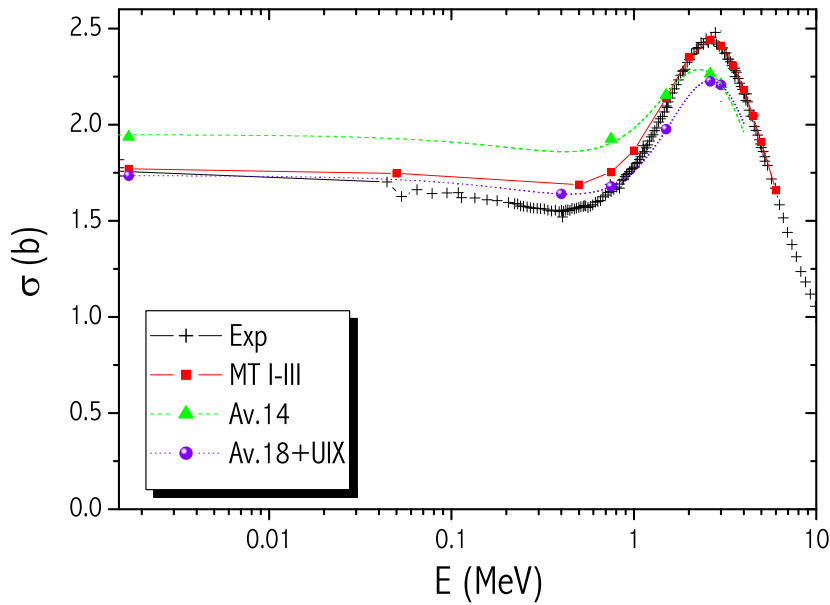}
\end{center}
\par
\vspace{-1.2cm} \caption{Calculated n+$^{3}$H total cross sections
compared with experimental data \protect\cite{Phlips}.}
 \label{Fig_nt_total_cs}
\end{figure}

Semi-realistic MT I-III potential was shown to be very successful
in describing total as well as differential cross sections \cite
{Carbonell} at the energies bellow n-n-d threshold. Situation is
less obvious for realistic potentials, which require much larger
partial wave basis (PWB) to obtain converged results. Recently we
have managed to considerably enlarge our PWB compared to
\cite{Carbonell1}. Despite of having some effect on the negative
parity phase shifts, the total cross section near the resonance
peak ($E_{cm}=3$ $MeV$) has not been improved (see Fig. \ref
{Fig_nt_total_cs}). We should notice, however, that the
$\mathcal{J}^{\Pi }=2^{-}$ phase shifts, the most relevant due to
its statistical factor, are not fully converged yet. They still
show some trend to increase, however we estimate that it can not
suffice to reproduce the experimental data.

\begin{figure}[htbp]
\begin{minipage}[t]{64mm}
\vspace{-0.4cm}\begin{center}\epsfxsize=64mm\mbox{\epsffile{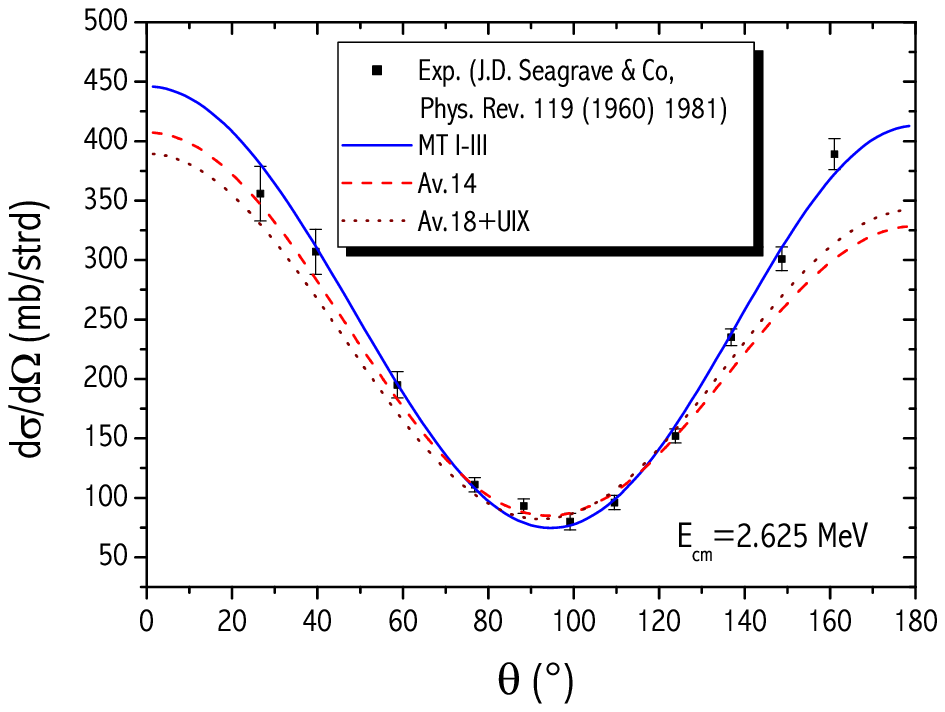}}\end{center}
\end{minipage}
\hspace{0.2cm}
\begin{minipage}[t]{64mm}
\vspace{-0.4cm}\begin{center}\epsfxsize=62mm\mbox{\epsffile{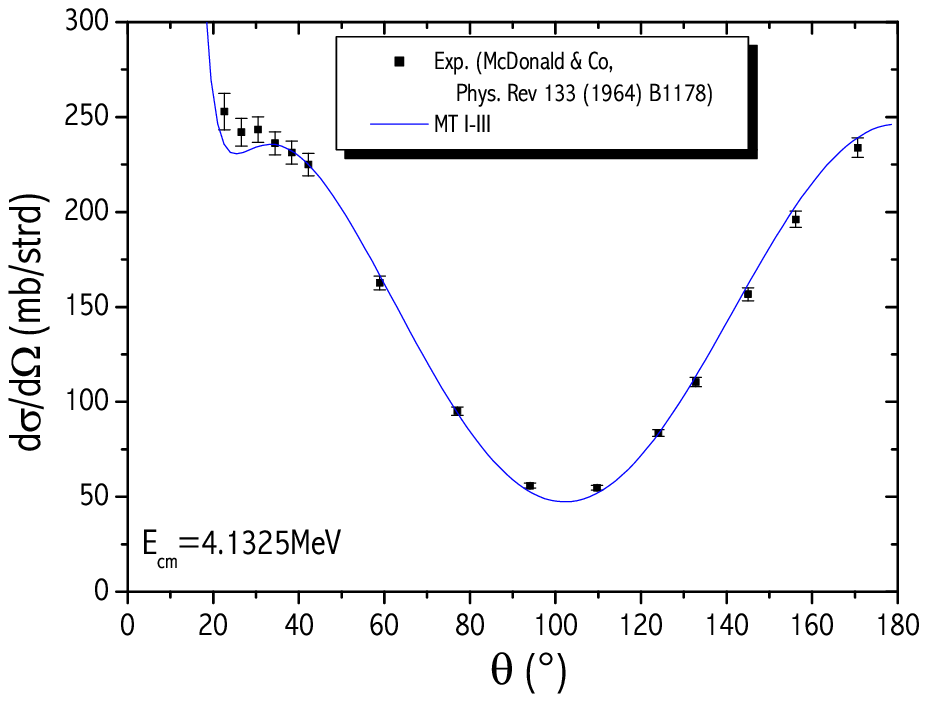}}\end{center}
\end{minipage}
 \vspace{-0.7cm} \caption{Differential cross
sections versus centre-of-mass angle for n+$^3$H at E$_{cm}=2.625$
MeV (figure on the left) and for p+$^3$He at E$_{cm}=4.1325 $ MeV
(figure on the right)} \label{diff_cs_phe}
\end{figure}

By including UIX-3NF we have managed to reproduce the experimental
zero-energy cross sections (see Table  \ref{table:1}), which are
overestimated by realistic NN interaction without 3NF. However
their effect near the peak remains very small. Other smaller
discrepancy exists near the minima of total cross sections
($E_{cm} \approx 0.4 MeV$). In this region, determined by positive
parity states, both MT I-III and Av.18+UIX overestimate
experimental data. This suggest that low energy parameters (0$^+$
and 1$^+$ scattering lengths and effective ranges) are not
described sufficiently well.

\subsection{p-$^3$He elastic scattering}
This system is an isospin partner of n-$^3$H. Yet calculations
were done uniquely with MT I-III model, except of
$\mathcal{J}^{\Pi }=0^{+}$ scattering lengths for which our Av.18
and Av.18+UIX predictions agree very well with the results of Pisa
group \cite{Pisa}. Like in the n-$^{3}$H case, MT I-III was found
to be successful in describing differential cross sections up to
d+p+p threshold (see Fig. \ref{diff_cs_phe}).

\subsection{p+$^{3}$H scattering at very low energies
\label{Sec_4N_pt}}

$^{4}$He continuum is the most complex 4N system, since its
spectrum contains numerous resonances (see Fig. \ref
{Fig_n4_en_chart}). Calculations of p+$^{3}$H scattering are
furthermore complicated by the existence of the first
$\mathcal{J}^{\Pi }=0^{+}$ excitation of $^{4}$He in its
thresholds vicinity. This resonance located at $E_{R}\approx0.4$
$MeV$ above p+$^{3}$H threshold covers with its width $\Gamma
\approx0.5$ $MeV$ almost the entire region below n+$^{3}$He. In
order to separate n+$^{3}$He and p+$^{3}$H channels one should
properly treat Coulomb interaction, the task is being furthermore
burden since both thresholds are described by the same isospin
quantum numbers.

When ignoring Coulomb interaction, as was a case in the most of
the calculations performed until now, n+$^{3}$He and p+$^{3}$H
thresholds coincide. $0^+$ resonant state moves below the joint
threshold and becomes a bound state for all NN potentials models
we use. Former fact is reflected in low energy scattering
observables: on one hand N+NNN scattering length in $0^+$ state is
found positive, on the other hand  excitation function decreases
smoothly with incident particles energy and does not show any
resonant behavior.

By properly taking Coulomb interaction into account, thus
separating n+$^{3}$He and p+$^{3}$H thresholds, we have placed the
$^4$He virtual state in between. However unlike in the other 4N
systems, MT I-III predictions for p-$^{3}$H singlet
($\mathcal{J}^{\Pi }=0^{+}$)
scattering length  as well as the excitation function -- $\left.{\frac{%
d\sigma}{d\Omega}}(E)\right|_{\theta=120^{\circ}}$ -- are in
disagreement
with experimental data. In  this model $^4$He virtual state is obtained too close to the p-$%
^{3}$H threshold and therefore has very small width. Nevertheless
out of resonance region the excitation function approaches
experimental data points, which is not a case in the calculations
without Coulomb interaction considered.
\begin{figure}[ht]
\begin{center}
\vspace{-0.25cm}\includegraphics[width=12.5cm]{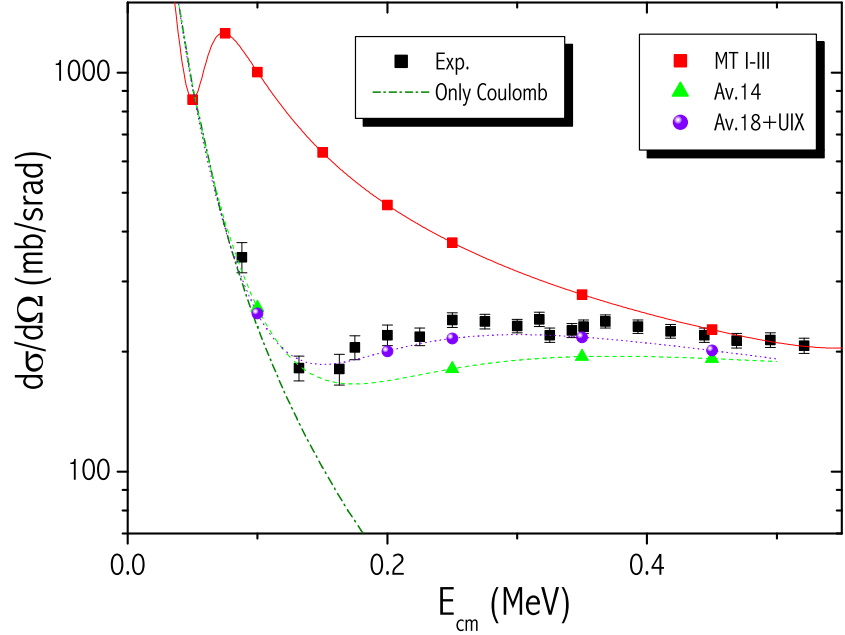}
\vspace{-1.5cm}
\end{center}
\caption{Energy dependence of p+$^{3}$He elastic differential
cross sections at 120$^{\circ}$: experimental results are compared
with our calculation.} \label{Fig_pt_cross_av}
\end{figure}
Our calculations with realistic potentials are still limited in
PWB. Nevertheless, they provide very promising results displayed
in Fig. \ref {Fig_pt_cross_av}. Pure 2NF models predict too large
singlet scattering length, thus placing the virtual state too far
from threshold. On the other hand, by implementing UIX-3NF in
conjunction with Av.18 NN model one
obtains singlet scattering length as well as the excitation function $\left.{%
\frac{d\sigma}{d\Omega}}(E)\right|_{\theta=120^{\circ}}$ in
agreement with experimental data \cite{Pt_exp}. A detailed
analysis of these calculations is in progress
\cite{Thesis,LC_JSPQ_03} as well as their extension above the
n-3He threshold.

\end{document}